\documentclass{article}
\usepackage{graphicx} 
\usepackage{amsmath, amsthm, amssymb}
\numberwithin{equation}{section}
\usepackage{hyperref}
\usepackage[left=1in, bottom=1in, right=1in, top=1in]{geometry}

\usepackage{authblk}

\usepackage{array}
\usepackage[authoryear, square]{natbib}
\usepackage{xcolor}
\usepackage{algpseudocode}
\usepackage{soul}
\usepackage{algorithm}
\usepackage{pdflscape}
\usepackage{soul}
\usepackage{makecell} 
\usepackage{ulem}
\usepackage{caption}

\usepackage{multirow}  

\usepackage{graphicx}
\usepackage{booktabs}
\newcommand{\indep}{\perp \!\!\! \perp}
\usepackage{placeins}   

\title{LARGE: A Locally Adaptive Regularization Approach for Estimating Gaussian Graphical Models}
\author[1]{Ha Nguyen \footnote{Email: tn325@cornell.edu}}
\author[1]{Sumanta Basu \footnote{Email: sumbose@cornell.edu}}
\affil[1]{Department of Statistics and Data Science, Cornell University}
\date{\today}

\begin{document}

\maketitle

\begin{abstract}
The graphical Lasso (GLASSO) is a widely used algorithm for learning high-dimensional undirected Gaussian graphical models (GGM). Given i.i.d. observations from a multivariate normal distribution, GLASSO estimates the precision matrix by maximizing the log-likelihood with an \( \ell_1 \)-penalty on the off-diagonal entries. 
However, selecting an optimal regularization parameter $\lambda$ in this unsupervised setting remains a significant challenge. A well-known issue is that existing methods, such as out-of-sample likelihood
maximization, select a single global $\lambda$ and do not account for heterogeneity in variable scaling or partial variances. Standardizing the data to unit variances, although a common workaround, has been shown to negatively affect graph recovery. 
Addressing the problem of nodewise adaptive tuning in graph estimation is crucial for applications like computational neuroscience, where brain networks are constructed from highly heterogeneous, region-specific fMRI data.

In this work, we develop Locally Adaptive Regularization for Graph Estimation (LARGE), an approach to adaptively learn nodewise tuning parameters to improve graph estimation and selection. In each block coordinate descent step of GLASSO, we augment the nodewise Lasso regression to jointly estimate the regression coefficients and error variance, which in turn guides the adaptive learning of nodewise penalties. We also employ a \textit{guiding} procedure that prioritizes pairs of nodes with high marginal correlations to promote early detection of true edges. Our method offers a conceptual shift in sparsity control: rather than relying on a scale-sensitive penalty $\lambda$, users specify a significance level $\alpha$, drawing on standard hypothesis testing for better interpretability. 
In simulations, LARGE consistently outperforms benchmark methods in graph recovery, demonstrates greater stability across replications, and achieves the best estimation accuracy in the most difficult simulation settings. We demonstrate the practical utility of our method by estimating brain functional connectivity from a real fMRI data set. 
\end{abstract}


\section{Introduction}\label{sec:intro}

Undirected Gaussian graphical models (GGMs) provide a useful framework for representing the network structure of multivariate data. Let  \( \mathcal{G} = (\mathcal{V}, \mathcal{E}) \) be an undirected graph where the vertex set \( \mathcal{V} = \{ 1, 2, \dots, p \} \) represents a $p$-dimensional random vector $X = [X_1, X_2, \ldots, X_p]^\top$, and the edge set $\mathcal{E}$ captures conditional dependence between nodes. The absence of an edge between two nodes $(j,k)$ in \( \mathcal{G} \) implies that the corresponding components $(X_j, X_k)$ are conditionally independent given all remaining variables. When $X$ follows a multivariate normal distribution $X \sim \mathcal{N}_p(0, \Sigma)$, this conditional independence structure is captured by the sparsity pattern of the precision matrix \( \Theta = \Sigma^{-1} \). Graph selection (structure learning) and estimation of the precision matrix is an important statistical task with applications in various domains, such as genetics, finance and social networks \citep{WANG2014gene, fan2011econ, COSTANTINI2019psych}.

In low-dimensional settings, the maximum likelihood estimator (MLE) of the precision matrix is simply the inverse of the sample covariance matrix. But in high-dimensional cases, the MLE is undefined when $p > n$, or may be unstable for large $p$. One of the earliest regularized methods for learning high-dimensional GGMs was developed by \citet{Meinshausen2006NR}, who proposed using nodewise regression (NWR) to recover the sparsity structure of the precision matrix. This method leverages the conditional independence structure of GGMs by fitting a Lasso regression for each node against all other nodes. Cross-validation (CV) is commonly used to select a separate tuning parameter for each regression, yielding $p$ different chosen values.

In contrast to NWR, the graphical Lasso (GLASSO) is a widely used method when both structure learning and estimation of the full precision matrix are of interest. It estimates the precision matrix by minimizing the negative log-likelihood with an added $\ell_1$-penalty on the off-diagonal entries of $\Theta$ \citep{yuan2007model, Banerjee2008}. GLASSO yields a sparse and interpretable precision matrix in high dimensions, and leads to a convex optimization problem that can be efficiently solved via block coordinate descent (BCD) \citep{friedman2008sparse}. A critical aspect of GLASSO implementation is the choice of the tuning parameter $\lambda$, which controls the level of sparsity, but tuning remains a challenging task in practice.

A common approach for selecting $\lambda$ is out-of-sample likelihood maximization or K-fold CV  \citep{friedman2008sparse}. This method splits the data into training and test sets, and uses grid search to find the $\lambda$ that yields the highest likelihood score on a held-out test set. Other grid search strategies have been proposed to select a tuning parameter that optimizes specific information-theoretic criteria. For example, the Risk Inflation Criterion (RIC) uses permutations to eliminate all associations between variables and selects the smallest $\lambda$ for which the estimated precision matrix is diagonal \citep{lysen2009ric}. Intuitively, this approach minimizes the chance of false positives in a truly empty graph. \cite{foygel2010extended} proposed selecting the $\lambda$ that optimizes the extended Bayesian Information Criterion (EBIC) to balance model fit and graph density. Alternatively, \cite{liu2010stars} proposed a subsampling-based method called Stability Approach to Regularization Selection (StARS). It partitions the data into $K$ subsets and selects the $\lambda$ that maximizes the stability of the estimated edge sets across subsamples.

A fundamental limitation of the existing approaches is that they apply the same penalization using a single $\lambda$ regardless of heterogeneity in the diagonal entries of the precision matrix. In GGMs, the diagonal entries of $\Theta$ encode the conditional precisions (inverse partial variances) of each variable given all others. Because these partial variances can differ substantially across variables, applying a uniform $\lambda$ to all entries ignores this difference in nodewise conditional variability. This issue has been well-documented in the graphical modeling literature \citep{sun_scaled_2013, Ren2015optimality, jankova2018inference, Carter2023partial}. 
We illustrate the cost of ignoring heterogeneous partial variances in Figure~\ref{fig:AR1_corrplot}: methods that rely on a single $\lambda$ tend to either under- or over-estimate different regions of in a simulated network. 
To handle heterogeneous tuning, the GLASSO algorithm (implemented in the R package \texttt{glasso}) allows for entry-wise penalties using a matrix whose \((j,k)\)-th element is \(\sqrt{\lambda_j \lambda_k}\), where \(\lambda = [\lambda_1, \lambda_2, \ldots, \lambda_p]^\top\) is a user-specified vector \citep{friedman2008sparse}.
 However, there is no universal guidance on how to appropriately choose these parameters.    

As a workaround, standardizing the observed data to have unit sample variances prior to graph estimation is often recommended. However, \citet{Carter2023partial} showed with simulations that standardization can in fact harm structure learning, in some cases even making it impossible to recover the underlying graphical model for any values of $\lambda$.
To address this, they proposed the Partial Correlation Graphical Lasso (PC-GLASSO) approach, which jointly estimates the diagonals of $\Theta$ and the partial correlations to produce an estimator of the full precision matrix. Their approach shifts the target of estimation from the precision matrix to the scale-invariant partial correlation matrix, using a global tuning parameter $\lambda$ chosen via the BIC criteria. A penalty $\mathrm{pen}_{ii} = 2 \log (\Theta_{ii})$ is applied to the diagonal elements to ensure scale invariance of the PC-GLASSO.

Addressing the heterogeneous tuning problem for GLASSO is crucial in real-world applications with heterogeneous noise levels and variable scaling. For example, in neuroscience applications, GLASSO is commonly used to estimate brain connectivity from functional Magnetic Resonance Imaging (fMRI) data \citep{Smith2011fmri, zhu_sparse_2018}. However, brain regions often exhibit region-specific noise characteristics, subject motion, and physiological variability \citep{ birn2014influence, liu2016noise}, all of which contribute to heterogeneous partial variances.
Relying on a single global tuning parameter when estimating functional connectivity can lead to systematic errors, missing true connections in some sub-networks while falsely identifying spurious edges in others. Such distortions undermine the validity of connectivity-based biomarkers, which are widely used to study cognitive performance and predict neurological disorders \citep{marek2022reproducible}.

In this paper, we propose Locally Adaptive Regularization for Graph Estimation (LARGE), an approach to adaptively learn nodewise tuning parameters that builds upon GLASSO to improve structure learning and estimation accuracy of GGMs. First, our approach leverages the well-known conceptual connection between GLASSO and NWR: within each BCD step, we augment the inner Lasso subproblems to also track the error variance of each regression. The $p$ noise variance estimators are then used to adaptively guide the selection of $p$ nodewise tuning pamaters. Second, we observe that without standardization to unit variances, GLASSO tend to favor edges with high marginal variances in the Lasso path, as thresholding operates on raw covariance. To address this, we introduce a \textit{guiding} procedure that promotes early detection of true edges by prioritizing pairs of nodes with high absolute marginal correlations.
Third, our algorithm performs sequential F-tests for edge selection with a user-defined significance threshold $\alpha$, replacing the scale-sensitive penalty $\lambda$. This reframing enables direct control over graph sparsity control through hypothesis testing and enhances interpretability for practitioners. Finally, our R package \texttt{large} is available \texttt{https://github.com/hanguyen97/large}, with computationally intensive functions written in C++. 

We conduct extensive numerical experiments on simulated data in both low- and high-dimensional settings to assess the performance of LARGE. First, our analysis confirms that using scale-adaptive tuning parameters improves graph selection performance over existing methods. This serves as a proof of concept that locally adaptive penalties enhance graph recovery compared to a single global $\lambda$ applied across all regressions. We also find that LARGE yields notably lower variability across replications, indicating improved stability. Second, while both LARGE and alternative benchmarks yield comparable performance when $n > p$, our method consistently outperforms them in the most challenging regime where $n = p$. Finally, we demonstrate the practical utility of LARGE by applying it to estimate functional connectivity from a real fMRI dataset in the Human Connectome Project (HCP).

The rest of the paper is organized as follows. In Section 2, we provide background on GLASSO and Autotune Lasso, the underlying procedure used to learn nodewise penalties. In Section 3, we introduce our proposed method, LARGE, and detail key design choices for the algorithm, including tuning initialization, adaptive regularization and sparsity control via significance testing. Sections 4 and 5 evaluate the performance of our method on simulated and real fMRI data, respectively.

\textbf{Notation.} Throughout the paper, for a vector \( \beta = [\beta_1, \beta_2, \ldots, \beta_p]^\top \in \mathbb{R}^p \), we write $\beta_{-j} \in \mathbb{R}^{p-1} $ for the vector obtained
by removing the $j$-th element. We use the notation 
\( \|\beta\|_q \) to denote the \( q \)-norm of \( \beta \), i.e., \( \left( \sum_i |\beta_i|^{q} \right)^{1/q} \) for any $q > 0$. A \( p \)-dimensional random vector is denoted by \( X = [X_1, X_2, \ldots, X_p]^\top \in \mathbb{R}^p \). Let \( X_1, X_2, \ldots, X_n \) be independent and identically distributed (i.i.d.) random samples from \( X \), where each sample \( X_i = [X_{i1}, X_{i2}, \ldots, X_{ip}]^\top \) for \( 1 \leq i \leq n \). We denote the resulting data matrix \( \mathcal{X} \in \mathbb{R}^{n \times p} \) as
\[
\mathcal{X} = 
\begin{bmatrix}
X_{11} & X_{12} & \cdots & X_{1p} \\
X_{21} & X_{22} & \cdots & X_{2p} \\
\vdots & \vdots & \ddots & \vdots \\
X_{n1} & X_{n2} & \cdots & X_{np}
\end{bmatrix}.
\]
For a matrix \(A \), we write \(A^\top\) for its transpose, \(A_{\cdot j}\) for its \(j\)-th column, and \(A_{\cdot,- j}\)  for the matrix obtained by removing the $j$-th column from $A$. 
Elements of matrices are denoted by lower indices, i.e., \(A_{ij}\). 
For two matrices \( A \) and \( B \), the Frobenius inner product between \( A \) and \( B \), denoted by \( \langle A, B \rangle \), is defined as \( \mathrm{tr}(A^\top B) \). If \(A\) is a square matrix, \(\mathrm{tr}(A)\) denotes its trace, i.e., the sum of its eigenvalues. The Frobenius norm of a matrix \(A\) is then given by $
\|A\|_F = \sqrt{\langle A, A \rangle}
$. We use \(\mathbb{S}_{+}^p\) to denote the cone of positive semi-definite \(p \times p\) matrices,
$
\mathbb{S}_{+}^p := \{ A \in \mathbb{R}^{p \times p} \mid A = A^\top, \ A \succeq 0 \},
$
and \(\mathbb{S}_{++}^p\) to denote the set of positive definite \(p \times p\) matrices,
$
\mathbb{S}_{++}^p := \{ A \in \mathbb{R}^{p \times p} \mid A = A^\top, \ A \succ 0 \} $.

\section{Background}\label{sec:background}





We begin with a brief introduction to the GGM framework. Let \( \mathcal{G} = (\mathcal{V}, \mathcal{E}) \) be an undirected graph with vertex set \( \mathcal{V} = \{1,2, \ldots, p\} \) and edge set \( \mathcal{E} \subset \mathcal{V} \times \mathcal{V} \). Consider a $p$-dimensional random vector \( X = [X_1,X_2, \ldots, X_p]^\top \), where each component is indexed by a vertex in the graph. The distribution of \( X \) is said to follow the underlying graphical model defined by \( \mathcal{G} \) if, for any pair of vertices \( j \) and \( k \), the corresponding variables \( X_j \) and \( X_k \) are conditionally independent given all other variables. 
Assume $
X \sim \mathcal{N}_p(0, \Sigma)
$, it can be shown that the inverse covariance matrix, or precision matrix $\Theta = \Sigma^{-1}$, encodes the conditional independence relationship among the elements of $X$ in the sparsity patterns in the off-diagonal elements. Specifically,
\begin{equation}
X_j \indep X_k \mid X_{\mathcal{V} \setminus \{(j,k)\}} \,\,\, \Leftrightarrow
\,\,\,
\Theta_{jk} = 0 \,\,\, \Leftrightarrow
\,\,\, (j,k) \notin \mathcal{E}. 
\end{equation}
The magnitudes of nonzero entries reflect the strength of conditional association. Consider i.i.d.\ samples \( X_1, \ldots, X_n \in \mathbb{R}^p \) of size $n$, our goal is to estimate the precision matrix and recover the edge set $\mathcal{E}$ of the underlying graph $\mathcal{G}$. This is also known as the \textit{(inverse) covariance selection} problem \citep{dempster1972covariance}.


\subsection{Precision Matrix Estimation with GLASSO}\label{subsec:GLASSO}

To solve this problem, the GLASSO estimator minimizes the $\ell_1$-penalized negative log-likelihood over the space of positive definite precision matrices:
\begin{equation}
    \hat{\Theta}_{\mathrm{GLASSO}} := 
    \arg\min_{\Theta \in \mathbb{S}_{++}^p} \left\{-  \log \det({\Theta}) + \mathrm{tr}({S} {\Theta}) + \lambda \|{\Theta}\|_{1, \text{off} } \right\},
    \label{eq:glasso_loglik} 
\end{equation}
where $S = \frac{1}{n} \sum_{i=1}^n X_i X_i^\top $ is the sample covariance matrix, 
$
\|\Theta\|_{1,\text{off}} = \sum_{i \ne j} |\Theta_{ij}|$, and \(\lambda > 0\) is a tuning parameter. The Karush-Kuhn-Tucker (KKT) conditions for this convex optimization problem are:
\begin{equation}
- {\Theta}^{-1} + {S} + \lambda \cdot {\Gamma}= 0
\text{ where }
\Gamma_{ij} = 
\begin{cases} 
\text{sign}(\Theta_{ij}) & \text{if } \Theta_{ij} \neq 0, \\
\in [-1, 1] & \text{if } \Theta_{ij} = 0.
\end{cases}
\label{eq:subgrad} 
\end{equation} \cite{friedman2008sparse} proposed a BCD algorithm to efficiently solve this system by iteratively updating one row and column of \( \Theta^{-1} \) while keeping the rest fixed. Let \( W \) denote the current estimate of \( \Theta^{-1} \). At each iteration, the target row and column are permuted to the last position, and \( W \) and \( S \) are partitioned as follows:
\[
W = \begin{bmatrix}
W_{11} & w_{12} \\
w_{12}^\top & w_{22}
\end{bmatrix}, \quad
S = \begin{bmatrix}
S_{11} & s_{12} \\
s_{12}^\top & s_{22}
\end{bmatrix}.
\]
where \(W_{11}, S_{11} \in \mathbb{R}^{(p-1)\times(p-1)}\), 
\(w_{12}, s_{12} \in \mathbb{R}^{p-1}\), 
and \(w_{22}, s_{22} \in \mathbb{R}\). 
Here, we use lowercase letters to denote the column vectors and matrix entries for clarity in the block partition. Using matrix partitioned inverses, the upper-right block of~\eqref{eq:subgrad} can be reformulated as  
\begin{equation}
    W_{11} \beta - s_{12} + \lambda \cdot \Gamma_{12} = 0, \quad \text{where } \beta = -\frac{\Theta_{12}}{\Theta_{22}}  \in \mathbb{R}^{p-1}.
    \label{eq:modified_lasso}
\end{equation}
Let $V=W_{11}$. This leads to a modified Lasso formulation that allows efficient updates of $\hat{\beta}$ via coordinate descent (CD):
\begin{equation}
    \hat{\beta}_k \leftarrow \mathcal{S}_{\lambda} \left( s_{12k} - \sum_{l \ne k} V_{kl} \hat{\beta}_l \right) \large/ V_{kk}; \quad k = 1, 2, \ldots, p-1,
    \label{eq:glasso_soft_threshold}
\end{equation}
 where \( \mathcal{S}_{\lambda}(z) = \text{sign}(z)(|z| - \lambda)_{+} \) denotes the soft-thresholding function. The GLASSO algorithm consists of an outer loop that updates one row and column of \( W \) at a time, and an inner loop that solves the modified Lasso problem ~\eqref{eq:modified_lasso} via coordinate-wise updates of \( \hat{\beta} \) until convergence.

\textit{\textbf{Connection between GLASSO and NWR}}. It is well-known that GLASSO is closely related to NWR \citep{hastie2015statistical}, which estimates the conditional dependencies of each node $j$ 
by regressing it on the remaining nodes \(\mathcal{V} \setminus \{j\} \) using Lasso. 
Let \(y = \mathcal{X}_{\cdot j}\) denote the \(j\)-th column of the data matrix \(\mathcal{X}\), 
and let \(\mathcal{Z} = \mathcal{X}_{ \cdot, - j }\) denote the matrix obtained by removing the \(j\)-th column. 
NWR estimates the sparse neighborhood of node \(j\) by solving:
\begin{equation}
\hat{\beta}^{\mathrm{NWR}} := \arg \min_{\beta \in \mathbb{R}^{p-1}} \left\{ \frac{1}{2n} (y - \mathcal{Z} \beta)^\top (y - \mathcal{Z}  \beta) + \lambda \| \beta \|_1 \right\},
\label{eq:lasso}
\end{equation}
where \(\|\beta\|_1 = \sum_{k=1}^{p-1} |\beta_k|\), and \(\lambda > 0\) is the tuning parameter. The corresponding subgradient condition is: 
\begin{equation}
\frac{1}{n} \mathcal{Z}^\top \mathcal{Z}  \beta - \frac{1}{n}\mathcal{Z} ^\top y + \lambda \, \text{sign}(\beta) = 0.
\label{eq:lasso_subg}
\end{equation}
The coordinate-wise update for \( \hat{\beta}^{\mathrm{NWR}} \) takes the form:
\begin{equation}
\hat{\beta}^{\mathrm{NWR}}_{k} \leftarrow  \mathcal{S}_{\lambda} \left( \frac{1}{n} \mathcal{Z}_{\cdot k}^\top y - \frac{1}{n} \sum_{l \ne k} \mathcal{Z}_{\cdot k}^\top \mathcal{Z}_{\cdot l} \hat{\beta}^{\mathrm{NWR}}_l \right) \Big/ \left(\frac{\| \mathcal{Z}_{\cdot k} \|^2_2}{n} \right) .
\end{equation}

Comparing with ~\eqref{eq:glasso_soft_threshold}, we observe that
$
s_{12,k} $ is the analog of $ \frac{1}{n} Z_{\cdot k}^\top y$. Similarly, $\sum_{\ell \ne k} V_{k\ell} \hat{\beta}_\ell $ and $ \frac{1}{n} \sum_{\ell \ne k} \mathcal{Z}_{\cdot k}^\top \mathcal{Z}_{\cdot \ell} \hat{\beta}_{\ell}^{\text{NWR}}$ both estimate $W_{11}$. As highlighted in \cite{friedman2008sparse}, GLASSO iteratively recomputes NWR regressions at each BCD step to update the precision matrix. This connection also underscores a key limitation: GLASSO effectively performs \( p \) separate regressions under-the-hood, yet applies a single global tuning parameter \( \lambda \) across all of them. This uniform penalization overlooks heterogeneity in partial variances, as reflected in the diagonal entries of $\Theta$. A more adaptive approach would assign nodewise tuning parameters $\lambda_j$ at each block update. However, finding multiple $\lambda_j$ via grid search would be computationally expensive and challenging due to the lack of a clear joint selection strategy. 

\textit{\textbf{Why data standardization is not sufficient.}} Running GLASSO on standardized data corresponds to solving:
\begin{equation}
\hat{\Theta}_{\mathrm{norm}} = \arg\min_{\Theta \in \mathbb{R}_{++}^p} \left\{ - \log \det(\Theta) +\mathrm{tr}(R \Theta)  + \lambda \|\Theta\|_{1, \mathrm{off}} \right\},
\end{equation}
where \( \hat{\Theta}_{\mathrm{norm}} \) estimates the inverse correlation matrix, \( D = \mathrm{diag}(S) \), and \( R = D^{-1/2} S D^{-1/2} \) is the sample correlation matrix. \( \hat{\Theta}_{\mathrm{norm}} \)  is of independent interest from the precision matrix, and is closely related to the weighted GLASSO estimator:
\begin{equation}
\hat{\Theta}_{\mathrm{w}} = \arg\min_{\Theta \in \mathbb{R}_{++}^p} \{ - \log \det(\Theta) + \mathrm{tr}(S \Theta) + \lambda \sum_{i \ne j} D^{1/2}_{ii} D^{1/2}_{jj} \left| \Theta_{ij} \right| \},
\end{equation}
where \( \hat{\Theta}_{\mathrm{norm}} = D^{1/2} \hat{\Theta}_{\mathrm{w}} D^{1/2} \) \citep{jankova2018inference}. The diagonals of \(\hat{\Theta}_{\mathrm{norm}}\), denoted 
\(D_{\mathrm{norm}} = \operatorname{diag}(\hat{\Theta}_{\mathrm{norm}})\), 
are not guaranteed to be standardized or uniform; in fact,
$
(D_{\mathrm{norm}})_{ii} 
= (\hat{\Theta}_{\mathrm{w}})_{ii} \,D_{ii}
$. As a result, applying a single global tuning parameter $\lambda$ across all nodes is not ideal.

This challenge reinforces the need for $p$ separate tuning parameters $\lambda_j$ when estimating $\Theta$. To address this, we adopt the Autotune Lasso procedure introduced by \cite{sadhukhan2025autotune} for each Lasso subproblem within the GLASSO algorithm. Autotune Lasso tracks an estimate of the noise variance and uses it to adaptively determine the tuning parameter in a regression problem. This approach has demonstrated promising results in sparse regression models, and we describe it in the next section. 

\subsection{Automatic Tuning via Autotune Lasso}\label{subsec:autotune_lasso}

Consider the linear model $y = \mathcal{Z}\beta + \varepsilon$ where $\varepsilon_i \overset{\mathrm{i.i.d.}}{\sim} \mathcal{N}(0, \sigma^2)$ from the NWR problem ~\eqref{eq:lasso}. Let each column of $\mathcal{Z}$ be standardized to have unit variance. Autotune Lasso jointly estimates the regression coefficients \( \beta \) and the error variance \( \sigma^2 \) via optimizing the full likelihood with an $\ell_1$-penalty: 
\begin{equation} 
(\hat{\beta}, \hat{\sigma}^{-2}) := \arg\min_{(\beta,  \sigma^{-2}) \in \mathbb{R}^{p-1} \times  \mathbb{R}_{+} } \left\{ 
 \frac{1}{2} \log(\sigma^{2}) + \frac{1}{2n\sigma^{2}}  \|y - \mathcal{Z}\beta\|_2^2 + \lambda_0 \|\beta\|_1 
\right\},
\label{eq:autotune_obj} 
\end{equation}
where $ \lambda_0 = \frac{1}{\mathrm{Var}(y)} \|\frac{1}{2n} \mathcal{Z}^\top y \|_{\infty}   $ is fixed. Computing the partial derivatives with respect to (w.r.t.) $\beta$ and $\sigma^{-2}$ to obtain the following closed-form updates:
 \begin{align}
\hat{\sigma}^{2} &\gets \frac{\|y - \mathcal{Z} \hat{ \beta }\|^2_2}{n } ,\label{eq:sigma2_starter} \\
    \hat{\beta}_k &\gets \mathcal{S}_{\lambda_0 \hat{\sigma}^{2}}   \left( \frac{1}{n} \mathcal{Z}_{\cdot k}^\top y - \frac{1}{n} \sum_{l \ne k} \mathcal{Z}^\top_{\cdot k} \mathcal{Z}_{\cdot l} \hat{\beta}_l \right) \nonumber  \\
    &= \mathcal{S}_{\lambda_0 \hat{\sigma}^{2}}   \left( \frac{1}{n} \langle \mathcal{Z}_{\cdot k}  , r_{k}\rangle \right)  
\end{align}
where $r_k = y - \sum_{l \ne k} \mathcal{Z}_{\cdot l } \hat{\beta}_l$ is the partial residual. Since the objective function ~\eqref{eq:autotune_obj} is biconvex in $\beta$ and $\sigma^{-2}$, \cite{sadhukhan2025autotune} adopts an alternating estimation approach that updates one set of parameters at a time while keeping the other fixed. For any given $\sigma^2$, solving for $\beta$ is essentially solving Lasso with tuning parameter $\lambda := \lambda_0 \hat{\sigma}^{2}$. Consequently, iteratively updating $\hat{\sigma}^{2}$ leads to solving a sequence of Lasso problems on an automatic, data-driven sequence of $\lambda$ values.

To improve the finite-sample estimation of \( \sigma^2 \), \cite{sadhukhan2025autotune} leverages the mean squared error (MSE) from a carefully constructed linear model as its estimator. At each iteration, Autotune Lasso ranks the importance of the regression coefficients \( \hat{\beta}_k \) based on the standard deviation (SD) of their partial residuals \( r_k \). Using forward selection guided by sequential F-tests, Autotune Lasso fits a linear model on a subset of predictors, and estimates $\sigma^2$ using the MSE:
\begin{equation}
    \hat{\sigma}^2 = \frac{\sum_{i=1}^n (y_i - \hat{y}_i)^2}{n - |\hat{S}|},
\end{equation}
where \( |\hat{S}| \) denotes the number of selected variables. For completeness, we present a simplified version of the Autotune Lasso procedure in Algorithm~\ref{alg:autotune_lasso}.




\begin{algorithm}[ht]
\caption{\textbf{Autotune Lasso \citep{sadhukhan2025autotune}} 
}
\label{alg:autotune_lasso}
\begin{algorithmic}[1]
\Statex \textbf{Input:} $y$, Normalized predictors $\mathcal{Z}$ (i.e., $\|\mathcal{Z}_{\cdot,j}\|_2^2 = n$), Sequential F-test significance level $\alpha$
\State Initialize $\hat{\beta} \gets 0$, $r \gets y$, $\hat{\sigma}^2 \gets \mathrm{Var}(y)$, $\lambda_0 \gets \frac{1}{\mathrm{Var}(y)} \left\|\frac{\mathcal{Z}^\top y}{2n} \right\|_\infty$, $\texttt{Predictor.Ranking} \gets \{1, \ldots, p-1\}$, $\texttt{Support.Set} \gets \emptyset$, $\texttt{sigma.update.flag} \gets \texttt{TRUE}$
\While{$\texttt{error} \geq \texttt{error.tolerance}$}
    \State $\texttt{Support.Set}^{(\text{old})} \gets \texttt{Support.Set},\quad \hat{\beta}^{(\text{old})} \gets \hat{\beta}$
    \For{$k$ in  \texttt{Predictor.Ranking}} \Comment{Estimate $\beta$}
        \State $\hat{\beta}_k \gets \texttt{Soft.Threshold}_{\lambda = \lambda_0 \hat{\sigma}^2}\left( \frac{1}{n} \mathcal{Z}_{\cdot k}^\top r + \hat{\beta}_k \right)$
        \State $r \gets r + \mathcal{Z}_{\cdot k} \left( \hat{\beta}_k^{(\text{old})} - \hat{\beta}_k \right)$
    \EndFor
    \If{$\texttt{sigma.update.flag} == \texttt{TRUE}$} \Comment{Estimate $\sigma^2$}
        \State $\texttt{Predictor.Ranking} \gets$ Rank $k$ by sorting $\mathrm{SD}(r_1), \ldots, \mathrm{SD}(r_{p-1})$ in decreasing order
        \State Fit an OLS model using sequential F-tests for variable selection with cutoff $=F_{\alpha; 1, n-i}$
\State \(\texttt{Support.Set} \gets  \{ k : \mathcal{X}_{\cdot k} \text{ selected into OLS} \}\)
\State $\hat{\sigma}^2 \gets  \frac{\sum_{i=1}^n \left( y_i - \hat{y}_i \right)^2 }{n - |\texttt{Support.Set}|}$
        \If{$\texttt{Support.Set} \subseteq \texttt{Support.Set}^{(\text{old})}$}
            \State $\texttt{sigma.update.flag} \gets \texttt{FALSE}$ \Comment{Early stopping of $\hat{\sigma}^2$ update}
        \EndIf
    \EndIf
    \State $\texttt{error} \gets \|\hat{\beta} - \hat{\beta}^{(\text{old})}\|_1 / \|\hat{\beta}^{(\text{old})}\|_1$
\EndWhile
\Statex \textbf{Output:} $\hat{\beta}, \hat{\sigma}^2, \lambda$
\end{algorithmic}
\end{algorithm}

\section{Method}\label{sec: method}

In this section, we demonstrate our method, Locally Adaptive Regularization for Graph Estimation. We extend the GLASSO estimator to allow for column-wise tuning parameters.
\begin{equation}
    \hat{\Theta}_{\mathrm{LARGE}} := 
    \arg\min_{\Theta \in \mathbb{S}_{++}^p} \left\{-  \log \det({\Theta}) + \mathrm{tr}({S} {\Theta}) + \sum_{j=1}^p \lambda_j \|{\Theta_{.j}}\|_{1,\mathrm{off}} \right\},
\label{eq:large_model} 
\end{equation}
where ${\|\Theta_{.j}}\|_{1,\mathrm{off}} = \sum_{k\ne j} |\Theta_{kj}|$. In the following, we describe step-by-step how the objective function~\eqref{eq:large_model} is solved using the \textsc{LARGE} algorithm. We outline the complete procedure in Algorithm~\ref{alg:large_alg}. We present an alternative formulation of LARGE, which includes penalization of both the diagonal and off-diagonal entries of $\Theta$ in the Appendix.

\begin{algorithm}[ht]
\caption{\textbf{Locally Adaptive Regularization for Graph Estimation (LARGE)}}
\label{alg:large_alg}
\begin{algorithmic}[1]
\Statex \textbf{Input:} Data matrix \(\mathcal{X} \in \mathbb{R}^{n \times p}\), Sequential F-test significance level $\alpha$
\State For $j=1,\ldots, p$, initialize \(\hat{\sigma}^2_j \gets \mathrm{Var}(\mathcal{X}_{\cdot j})\),  \(\lambda_{0j} \gets \frac{1}{\mathrm{var}(\mathcal{X}_{\cdot j})} \max_{k \ne j} \left| \frac{1}{2n} \langle \mathcal{X}_{\cdot j}, \mathcal{X}_{\cdot k} \rangle \right|\)
\State Let $S \gets \mathrm{Cov}(\mathcal{X})$; initialize $W \gets S$

\Repeat \text{ for \( j = 1, 2, \ldots, p \):}
\State (a) Partition $
W = \begin{bmatrix}
W_{11} & w_{12} \\
w_{12}^\top & w_{22}
\end{bmatrix}, \quad
S = \begin{bmatrix}
S_{11} & s_{12} \\
s_{12}^\top & s_{22}
\end{bmatrix}.
$ 

    \State (b) Solve $W_{11} \beta - s_{12} + \lambda_{j} \cdot \text{sign}(\beta) = 0$ using Autotune Lasso:
    \State Let $V \gets W_{11}$; initialize $\hat{\beta} \gets 0$
    \State Initialize $\texttt{Predictor.Ranking}_{j} \gets \{1, \ldots, p-1\}$, $\texttt{Support.Set}_{j} \gets \emptyset$, $\texttt{sigma.update.flag}_{j}  \gets \texttt{TRUE}$


    \While{$\texttt{error} \geq \texttt{error.tolerance}$}
    \State $\texttt{Support.Set}_j^{(\text{old})} \gets \texttt{Support.Set}_j,\quad \hat{\beta}^{(\text{old})} \gets \hat{\beta}$

        \For{$k$ in $\texttt{Predictor.Ranking}_{j}$} \Comment{Estimate $\beta$}
            \State \(\hat{\beta}_k \gets \texttt{Soft.Threshold}_{\textcolor{black}{\lambda_j = \hat{\sigma}^2_j \cdot \lambda_{0j}}} \left(s_{12k} - \sum_{l \ne k} V_{kl} \cdot \hat{\beta}_l \right) \big/ V_{kk}\) 
        \EndFor
        \If{$\texttt{sigma.update.flag}_j == \texttt{TRUE}$} \Comment{Estimate $\sigma^2_j$}
        \State $\texttt{Predictor.Ranking}_j \gets$ Rank \(k\) in decreasing order of:
        \[
        \begin{cases}
        \textcolor{black}{\left| \mathrm{Cor}(\mathcal{X}_{\cdot j}, \mathcal{X}_{\cdot k}) \right|}, & \text{if first iteration of line 8} \\
        \mathrm{SD} \text{ of } r_k = \mathcal{X}_{\cdot j} - \sum_{l \notin \{ j, k \}} \mathcal{X}_{\cdot l} \hat{\beta}_l, & \text{otherwise}
        \end{cases}
        \]
        \State Fit a linear model using sequential F-tests for variable selection with cutoff $=F_{\alpha; 1, n-i}$
        \State \(\texttt{Support.Set}_j \gets  \{ k : \mathcal{X}_{\cdot k} \text{ selected into the linear model} \}\)
        \State \textcolor{black}{\(\hat{\sigma}^2_j \gets \frac{\sum_{i=1}^n (X_{ij} - \hat{X}_{ij})^2}{n - |\texttt{Support.Set}_j|}\)}
        \If{$\texttt{Support.Set}_j \subseteq \texttt{Support.Set}_j^{(\text{old})}$}
            \State $\texttt{sigma.update.flag}_j \gets \texttt{FALSE}$
        \EndIf
     \EndIf
    \EndWhile
    \State (c) \( w_{12} \gets W_{11} \hat{\beta} \) with symmetrization
\Until{convergence over all \(j\)}
\State For $i = 1, \ldots, p$, \(\hat{\Theta}_{jj} \gets \hat{\sigma}_j^{-2}\) and \(\hat{\Theta}_{j,-j} \gets -\hat{\beta} \hat{\Theta}_{jj}\) with symmetrization
\Statex \textbf{Output:} \(\hat{\Theta}\), $(\hat{\sigma}_j^2, \lambda_j)$ for $j=1,\ldots, p$
\end{algorithmic}
\end{algorithm}

\textit{\textbf{Initializing nodewise tuning parameters.}} To initialize the optimization, we assign nodewise tuning parameter $\lambda_j$ for each component $\mathcal{X}_{\cdot j}$. Following \cite{sadhukhan2025autotune}, we let
$
\lambda_{0j} = \frac{1}{\mathrm{Var}(\mathcal{X}_{\cdot j})}\max_{k \ne j} \left| \frac{1}{2n} \langle \mathcal{X}_{\cdot j}, \mathcal{X}_{\cdot k} \rangle \right|
$, and initialize the noise variance estimate as 
 $\hat{\sigma}^2_j = \mathrm{Var}(\mathcal{X}_{\cdot j})$.
Substituting into the expression for $\lambda_{j}$, the starting tuning value for the Lasso subproblem of node $j$ becomes $\lambda_j =   \max_{k \ne j} \left| \frac{1}{2n} \langle \mathcal{X}_{\cdot j}, \mathcal{X}_{\cdot k}  \rangle\right|$ .

\textit{\textbf{Replacing the Lasso subproblem with Autotune Lasso.}}  
At the core of our algorithm, we follow the GLASSO framework, and replace the nodewise Lasso subproblem with an Autotune Lasso regression that adaptively learns the tuning parameter \( \lambda_j \) for each component $\mathcal{X}_{\cdot j}$. This substitution is the main distinction from GLASSO: instead of using a global penalty, we update \( \lambda_j \) based on the local noise level and use it to threshold \( \beta \). At each BCD step, we reformulate the upper-right block of the KKT condition~\eqref{eq:subgrad} as  
\begin{equation}
    W_{11} \beta - s_{12} + \lambda_j \cdot \Gamma_{12} = 0, \quad \text{where } \beta = - \frac{\Theta_{12}}{\Theta_{22}}   \text{ and } \lambda_j = \hat{\sigma}^2_j \lambda_{0j}.
    \label{eq:modified_atlasso}
\end{equation}
This leads to an alternating minimization scheme where \( \hat{\beta} \) is updated via CD:
\begin{equation}
    \hat{\beta}_k \leftarrow \mathcal{S}_{\lambda_j = \hat{\sigma}^2_j \lambda_{0j}} \left( s_{12k} - \sum_{l \ne k} V_{kl} \hat{\beta}_l \right) \large/ V_{kk},
    \label{eq:glasso_soft_threshold}
\end{equation} 
and the nodewise noise variance \( \hat{\sigma}^2_j \) is updated using the MSE of a carefully constructed linear model:
\begin{equation}
    \hat{\sigma}^2_j \gets \frac{\sum_{i=1}^n \left( X_{ij} - \hat{X}_{ij}\right)^2 }{n - |\mathit{\hat{S}}_j|},
    \label{eq:sigma2_hat_est}
\end{equation}
where $|\mathit{\hat{S}}_j|$ is the number of selected edges. The linear model is constructed via forward selection with sequential F-tests, as described in Subsection~\ref{subsec:autotune_lasso}. We then locally update the penalty level for node \( j \) as
$
\lambda_j = \lambda_{0j} \hat{\sigma}^2_j,
$
tuning the regularization strength based on the noise level for its corresponding Lasso subproblem.

The LARGE algorithm retains the outer loop of GLASSO, updating one row and column of \( W \) at a time, while the inner loop runs Autotune Lasso, until convergence. For BCD steps beyond the first (line 3 of Algorithm~\ref{alg:large_alg}), we warm-start \( \lambda_j \) using its value from the previous outer iteration instead of reinitializing at $\lambda_j =   \max_{k \ne j} \left| \frac{1}{2n} \langle \mathcal{X}_{\cdot j}, \mathcal{X}_{\cdot k}  \rangle\right|$. This preserves information across iterations and helps BCD get unstuck along the regularization path.

\textit{\textbf{Guiding the initial Lasso path.} } From Equation~\eqref{eq:glasso_soft_threshold}, we see that each coefficient \( \hat{\beta}_k \) is updated by thresholding the raw covariances. When the variances of the individual variables differ substantially, these covariances are dominated by features with large marginal variance. This helps explain the common recommendation to standardize variables before applying Lasso or GLASSO \citep{hastie2015chapter2}. In our setting, high-variance edges connected to a target node \( j \) tend to produce larger partial residuals, leading to the early inclusion of spurious edges along the Lasso path.

To address this, we \textit{guide} the initial path by ranking edge importance using absolute marginal correlations during the first iteration of the inner loop (line 8 in Algorithm~\ref{alg:large_alg}), before performing variable selection for a linear model using sequential F-tests. Specifically, for each target node \( j \), we rank the importance of \( \hat{\beta}_k \) based on its absolute marginal correlation:
\begin{equation}
    \left| \mathrm{Cor}(\mathcal{X}_{\cdot j}, \mathcal{X}_{\cdot k}) \right| =  \frac{\left|\langle \mathcal{X}_{\cdot j}, \mathcal{X}_{\cdot k} \rangle \right|}{\|\mathcal{X}_{\cdot j}\|_2 \, \|\mathcal{X}_{\cdot k}\|_2} .
\end{equation}
This step is motivated by the Sure Independence Screening (SIS) rule of \cite{fan2008sure}, which ranks variables by marginal correlation to reduce dimensionality before model fitting. Our goal is that by prioritizing edges with high marginal correlation with a target node $j$ at the outset, we obtain a more accurate initial linear model and estimate of $\hat{\sigma}_j^2$. 

We run a simple simulation using synthetic data to understand how the \textit{guiding} procedure affects the accuracy of noise variance estimates $\hat{\sigma}_j^2$. We generate i.i.d. data from $X \sim \mathcal{N}(0, \Theta^{-1})$ with $(n, p) = (200, 50)$. To simulate heterogeneous noise levels, we construct a tridiagonal matrix $\Theta \in \mathbb{R}^{50 \times 50}$ with two distinct sub-blocks: \(\Theta_{ii} 
 = 10, \Theta_{i,i+1} = \Theta_{i+1,i} = 3 \) for $i=1,\ldots, 25$, and \( \Theta_{ii} 
 = 1, \Theta_{i,i+1} = \Theta_{i+1,i}   = 0.3 \) for $i=26,\ldots, 50$. We run LARGE with an F-test significance level of \( \alpha = 2\% \), repeating the experiment over 20 replications. For evaluation, we compare the estimated noise variance \( \hat{\sigma}_j^2 \) for a randomly selected node to the corresponding estimate from an oracle linear model.

\begin{figure}[ht]
    \centering
    \includegraphics[width=0.7\textwidth]{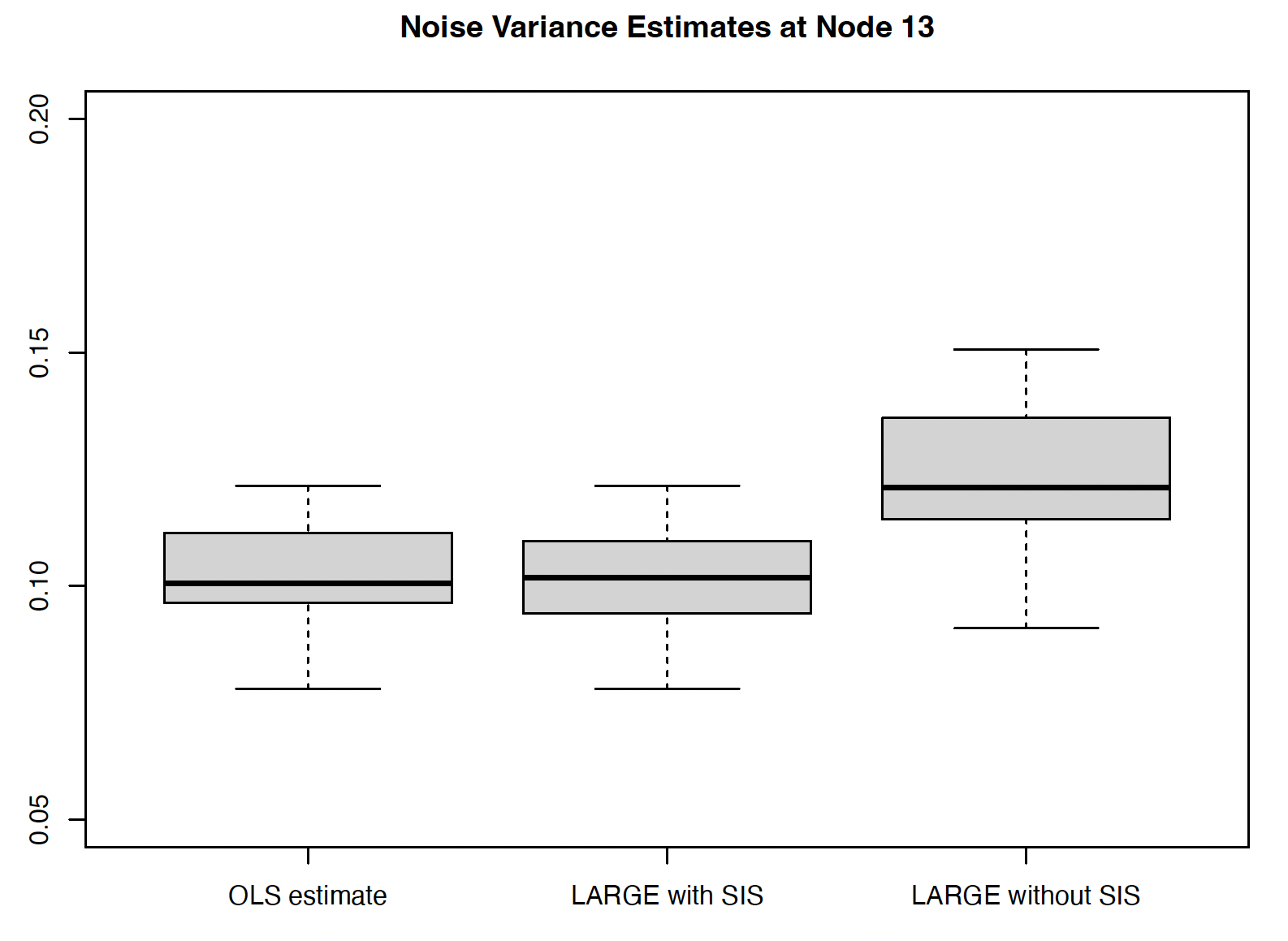}
    \caption{\textit{Boxplot of noise variance estimates \( \hat{\sigma}_j^2 \) for a randomly selected node under a tridiagonal precision matrix. Data are generated from $X \stackrel{\text{iid}}{\sim} \mathcal{N}(0, \Theta^{-1})$ with \( (n, p) = (200, 50) \), aggregated over 20 replications.} Estimates from the \textit{guiding} procedure more closely align with those from an oracle linear model.}
    \label{fig:sigma_convergence}
\end{figure}

As shown in Figure~\ref{fig:sigma_convergence}, the estimates of \( \hat{\sigma}_j^2 \) without the \textit{guiding} procedure are consistently inflated relative to the oracle linear benchmark. In contrast, after applying \textit{guiding}, the estimates align much more closely with the linear model reference, indicating improved convergence toward the true error variance.

\textit{\textbf{Controlling sparsity via significance testing.}} For each node, we use sequential F-tests, a classical stepwise regression tool, to perform variable selection and construct the linear model used to estimate $\sigma^2_j$. This procedure examines a sequence of nested linear models and their corresponding residual sum of squares (RSS) as more variables (edges) are added, and stops when the reduction in RSS is no longer statistically significant at the chosen significance level $\alpha$ \citep{Montgomery2020_ISQC}. This allows $\alpha$ to serve as the sole tuning parameter, offering direct and interpretable control over graph sparsity, unlike methods that rely on a scale-sensitive penalty parameter $\lambda$. By leveraging the classical F-statistic, this framework bypasses variable scaling issues and allows practitioners to select $\alpha$ based on the context of applications.


\section{Numerical Experiments}\label{sec:simulation}

In this section, we conduct comprehensive simulation studies to evaluate the graph selection and estimation  accuracy of LARGE. The F-test significance level is set to $\alpha = 2\%$ based on empirical performance across a range of simulation settings. We find that while setting $\alpha = 5\%$ yields a similar true positive rate, it leads to a slightly higher false positive rate.

\textbf{Benchmark:} We benchmark our method against existing tuning parameter selection methods for GLASSO, including K-fold CV, RIC, EBIC, and StARS. We use the \texttt{CVglasso} package \citep{CVglasso} to run 5-fold CV and the \texttt{huge} package for RIC, EBIC, and StARS \citep{huge}. For a fair comparison, we use each method to select the tuning parameter and then apply the native function from the \texttt{glasso} package to estimate $\Theta$ \citep{glasso}. 

Additionally, we compare against two other methods: (1) PC-GLASSO \citep{Carter2023partial}, which also addresses the heterogeneous tuning problem by decomposing graph estimation into separate penalized estimations of the partial correlations and the diagonal entries of $\Theta$; and (2) Tuning-Insensitive Graph Estimation and Regression (TIGER) \citep{liu2017tiger}, a commonly used method with tuning-insensitive properties.

\textbf{Generative models:} Building on the setup from \citet{Ren2015optimality}, we construct \( p \times p \) precision matrices \( \Theta \) consisting of three equally sized blocks of diagonal values with a ratio of \( 1\!:\!1\!:\!1 \). The diagonal entries are set to \( (\alpha_1, \alpha_2, \alpha_3) = (10, 1, 0.5) \) for the three blocks, respectively. The setup reflects the fact that nodewise partial variances can differ substantially, making it unlikely that a single global tuning parameter will perform well. We consider the following data-generating processes (DGPs) for \( \Theta \):

\begin{enumerate}
    \item \textbf{DGP1:} Band-1 graph model. Within each block, entries follow \( \Theta_{i, i+1} = \Theta_{i+1,i} = 0.3 \alpha_k \) for \( k \in \{1, 2, 3\} \).
    \item \textbf{DGP2:} Band-2 graph model with \( \Theta_{i, i+1} = \Theta_{i+1,i}  = 0.3 \alpha_k \) and \( \Theta_{i, i+2} = \Theta_{i+2,i} = 0.2 \alpha_k \).
    \item \textbf{DGP3:} Block graph model with three blocks, where each block is a random graph with 1\% sparsity. Nonzero entries are set to \( \Theta_{ij} = 0.4 \alpha_k \).
    \item \textbf{DGP4:} Hub graph model with \( p/15 \) hub nodes. Off-diagonal entries are scaled by \( \alpha_k \).
    \item \textbf{DGP5:} Random graph model with 1\% sparsity. Nonzero entries are set to $0.4$, then scaled according to the diagonals \( \alpha_k \).
\end{enumerate}
For each setting, we draw random samples from a multivariate Gaussian distribution \(\mathcal{N}_p(0, \Theta^{-1})\) with \(n \in \{300, 500\}\) and \(p \in \{100, 300\}\). All experiments are replicated 50 times. 

\textbf{Performance metrics:} To assess the ability to recover the correct sparsity structure, we calculate the area under the receiver operating characteristic (AUROC) curve. To evaluate estimation accuracy, we compute the relative mean squared error (RMSE) between the estimated and true precision matrices. Specifically, we focus on the off-diagonal elements:
\begin{equation}
\text{RMSE}_{\text{off}} = \frac{\| \hat{\Theta}_{\text{off}} - \Theta_{\text{off}} \|^2_F}{\| \Theta_{\text{off}} \|^2_F},
\end{equation}
where $\Theta_{\text{off}}$ denotes the matrix $\Theta$ with diagonal entries set to zero. We restrict the RMSE calculation to off-diagonal elements, as they encode the conditional dependencies between variables and directly determine the graph estimation performance.  We report the mean and SD of each performance metric across 50 replications. 

\subsection{Graph selection}

\begin{figure}[ht]
    \centering
    \includegraphics[width=1\textwidth]{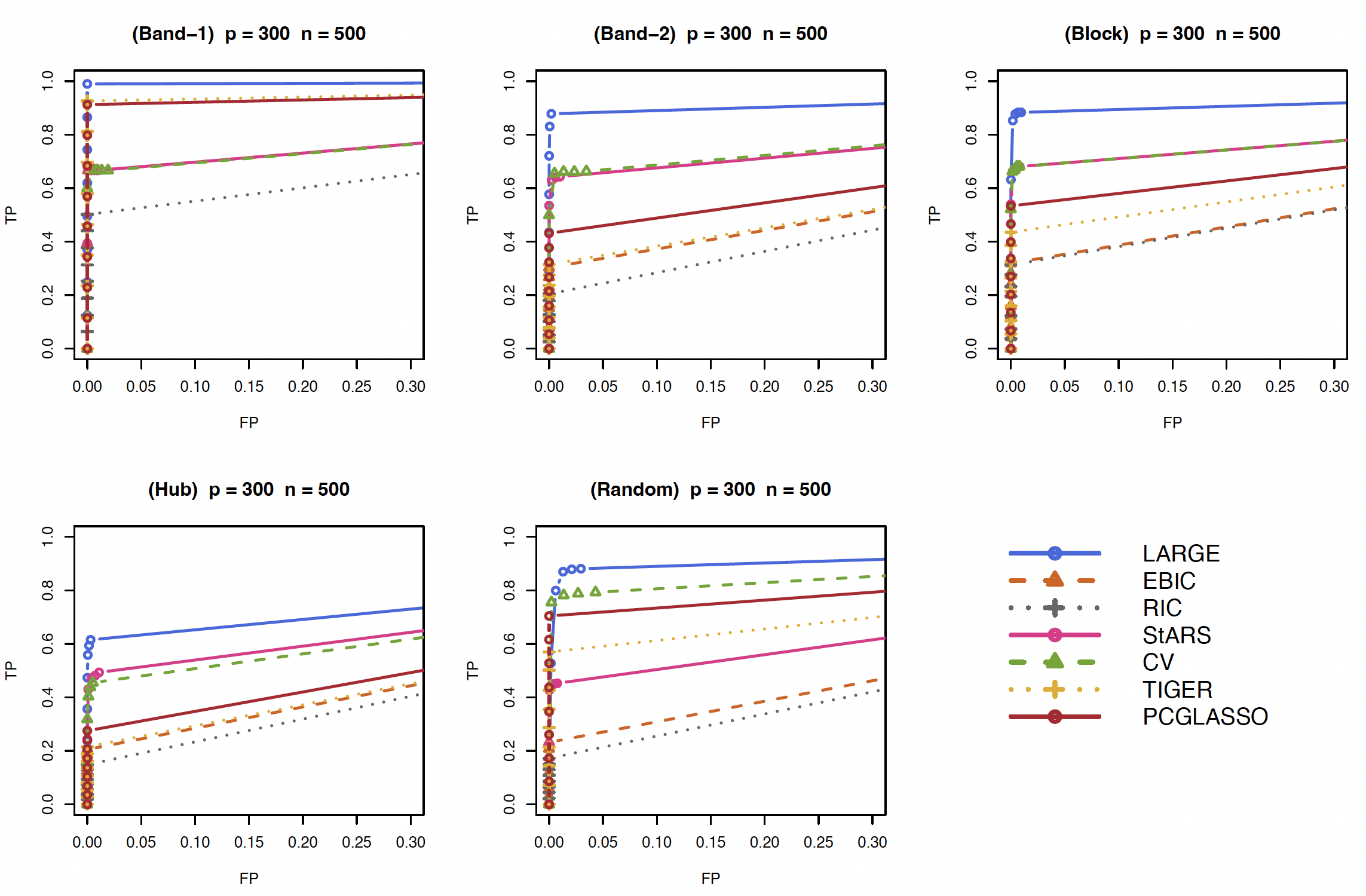}
    \caption{\textit{ROC curves from a single replication across all simulation settings where \( (n, p) = (500, 300) \).} In this high-dimensional regime, LARGE consistently achieves the highest AUROC across the various graph structures.}
    \label{fig:roc_curves}
\end{figure}

AUROC values are reported in Table~\ref{tab:auroc_results}, and example ROC curves are shown in Figure~\ref{fig:roc_curves}. LARGE achieves the highest AUROC across nearly all settings, indicating superior graph recovery under heterogeneous nodewise noise levels. Notably, LARGE also exhibits the lowest SD across replications, suggesting greater stability in performance. As expected, the CV method shows substantially higher SD in AUROC results, likely due to sensitivity to random train-test splits across replications. Interestingly, the performance of PC-GLASSO and TIGER drops sharply as the GDPs become more challenging, e.g., when the sample sizes reduce from $n=500$ to $n=300$ for the case $p=300$.

\begin{table}[!htbp]
\centering
\resizebox{\textwidth}{!}{%

\begin{tabular}{l|ccccccc}
\toprule
 &  &  & & AUROC &   &  & \\
Family & LARGE & EBIC & RIC & StARS & 5-fold CV & TIGER & PC-GLASSO \\
\midrule
Band-1 &  &  &  &  &  &  & \\
\hspace{1em}p = 100 &  &  &  &  &  &  & \\
\hspace{2em}n = 300 & 0.99 (0.01) & 0.75 (0.06) & 0.74 (0.02) & 0.84 (0) & 0.84 (0.02) & 0.97 (0.01) & \textbf{1 (0)} \\
\hspace{2em}n = 500 & \textbf{1 (0)} & 0.84 (0) & 0.82 (0.01) & 0.84 (0) & 0.90 (0.08) & \textbf{1 (0)} & \textbf{1 (0)} \\
\hspace{1em}p = 300 &  &  &  &  &  &  & \\
\hspace{2em}n = 300 & \textbf{0.98 (0.01)} & 0.70 (0.01) & 0.68 (0.01) & 0.83 (0) & 0.83 (0) & 0.93 (0.01) & 0.95 (0.01) \\
\hspace{2em}n = 500 & \textbf{1 (0)} & 0.83 (0) & 0.77 (0.02) & 0.83 (0) & 0.83 (0) & \textbf{1 (0)} & \textbf{1 (0)} \\
\midrule
Band-2 &  &  &  &  &  &  & \\
\hspace{1em}p = 100 &  &  &  &  &  &  & \\
\hspace{2em}n = 300 & \textbf{0.90 (0.01)} & 0.65 (0.01) & 0.61 (0.01) & 0.79 (0.02) & 0.83 (0.01) & 0.61 (0.01) & 0.75 (0.01) \\
\hspace{2em}n = 500 & 0.96 (0.01) & 0.67 (0.01) & 0.67 (0.01) & 0.81 (0.01) & 0.84 (0) & 0.76 (0.01) & \textbf{0.99 (0)} \\
\hspace{1em}p = 300 &  &  &  &  &  &  & \\
\hspace{2em}n = 300 & \textbf{0.86 (0.01)} & 0.63 (0.01) & 0.57 (0.01) & 0.80 (0.01) & 0.79 (0.02) & 0.55 (0) & 0.50 (0.01) \\
\hspace{2em}n = 500 & \textbf{0.94 (0.01)} & 0.66 (0.01) & 0.62 (0.01) & 0.82 (0) & 0.83 (0) & 0.69 (0.01) & 0.75 (0.01) \\
\midrule
Block &  &  &  &  &  &  & \\
\hspace{1em}p = 100 &  &  &  &  &  &  & \\
\hspace{2em}n = 300 & \textbf{0.87 (0.05)} & 0.66 (0.07) & 0.62 (0.06) & 0.78 (0.06) & 0.71 (0.11) & 0.61 (0.05) & 0.65 (0.15) \\
\hspace{2em}n = 500 & \textbf{0.92 (0.04)} & 0.67 (0.05) & 0.66 (0.05) & 0.79 (0.06) & 0.76 (0.1) & 0.81 (0.06) & 0.73 (0.12) \\
\hspace{1em}p = 300 &  &  &  &  &  &  & \\
\hspace{2em}n = 300 & \textbf{0.86 (0.02)} & 0.64 (0.02) & 0.59 (0.02) & 0.78 (0.03) & 0.73 (0.05) & 0.55 (0.01) & 0.50 (0) \\
\hspace{2em}n = 500 & \textbf{0.92 (0.01)} & 0.66 (0.02) & 0.65 (0.02) & 0.82 (0.02) & 0.81 (0.04) & 0.73 (0.03) & 0.80 (0.02) \\
\midrule
Hub &  &  &  &  &  &  & \\
\hspace{1em}p = 100 &  &  &  &  &  &  & \\
\hspace{2em}n = 300 & \textbf{0.80 (0.02)} & 0.60 (0.01) & 0.58 (0.01) & 0.71 (0.01) & 0.76 (0.03) & 0.56 (0.01) & 0.63 (0.05) \\
\hspace{2em}n = 500 & 0.84 (0.01) & 0.61 (0.01) & 0.62 (0.01) & 0.74 (0.01) & 0.79 (0.02) & 0.66 (0.01) & \textbf{0.87 (0.01)} \\
\hspace{1em}p = 300 &  &  &  &  &  &  & \\
\hspace{2em}n = 300 & \textbf{0.76 (0.01)} & 0.59 (0.01) & 0.54 (0.01) & 0.72 (0.01) & 0.71 (0.03) & 0.52 (0) & 0.50 (0) \\
\hspace{2em}n = 500 & \textbf{0.81 (0.01)} & 0.60 (0.01) & 0.59 (0.01) & 0.75 (0.01) & 0.76 (0.02) & 0.60 (0.01) & 0.64 (0.01) \\
\midrule
Random &  &  &  &  &  &  & \\
\hspace{1em}p = 100 &  &  &  &  &  &  & \\
\hspace{2em}n = 300 & \textbf{0.94 (0.02)} & 0.60 (0.03) & 0.58 (0.02) & 0.70 (0.03) & 0.70 (0.08) & 0.61 (0.03) & 0.74 (0.12) \\
\hspace{2em}n = 500 & \textbf{0.96 (0.02)} & 0.61 (0.03) & 0.64 (0.03) & 0.70 (0.03) & 0.77 (0.11) & 0.82 (0.03) & 0.80 (0.09) \\
\hspace{1em}p = 300 &  &  &  &  &  &  & \\
\hspace{2em}n = 300 & \textbf{0.93 (0.01)} & 0.61 (0.02) & 0.56 (0.01) & 0.72 (0.01) & 0.72 (0.03) & 0.61 (0.01) & 0.53 (0.07) \\
\hspace{2em}n = 500 & \textbf{0.95 (0.01)} & 0.63 (0.01) & 0.62 (0.01) & 0.73 (0.01) & 0.86 (0.06) & 0.83 (0.01) & 0.90 (0.01) \\
\bottomrule
\end{tabular} }
\caption{\textit{Mean  AUROC of LARGE and benchmark methods across all simulation settings.} Bolded values indicate the best performance in each setting. Standard
deviations are reported in parentheses. Results are averaged over 50 replications.}
\label{tab:auroc_results}
\end{table}

\begin{figure}[ht]
    \centering
    \includegraphics[width=1\textwidth]{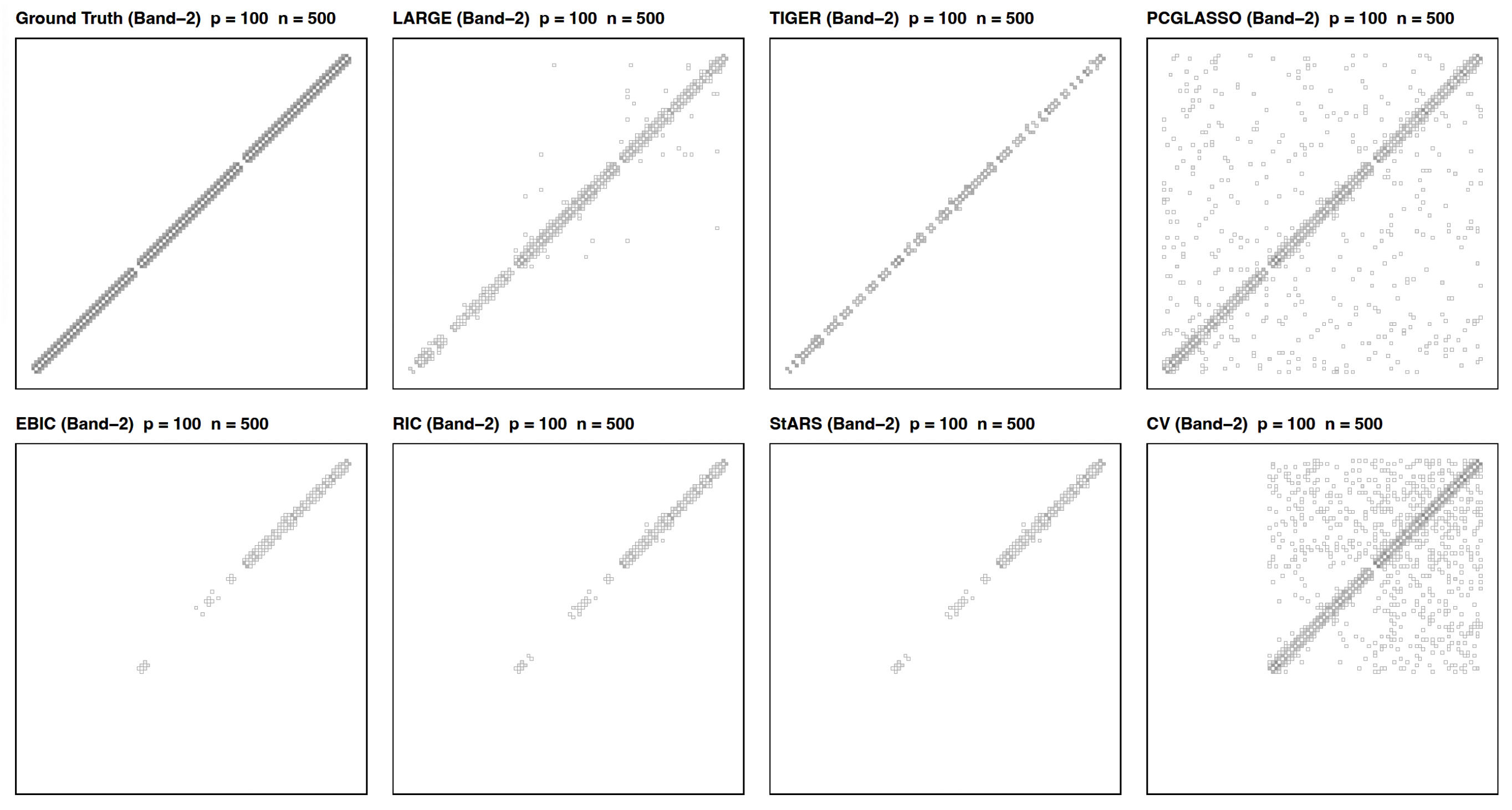}
    \caption{\textit{Estimated precision matrices $\hat{\Theta}$ for a band-2 graph model, plotted in absolute correlation scale. Data are generated from $X \stackrel{\text{iid}}{\sim} \mathcal{N}(0, \Theta^{-1})$ with \( (n, p) = (500, 100) \).} In the second row, methods relying on a global tuning parameter (EBIC, RIC, StARS, and CV) tend to either under- or over-estimate different regions of the network in the presence of heterogeneous local noise levels.}
    \label{fig:AR1_corrplot}
\end{figure}

To illustrate the importance of adapting the tuning parameter to the local noise levels, we visualize the estimated precision matrices from various methods for a band-2 graph model in Figure~\ref{fig:AR1_corrplot}. In this simple example, LARGE correctly identifies all true edges with minimal false positives. In contrast, EBIC, RIC, and StARS (second row of Figure~\ref{fig:AR1_corrplot}) select overly conservative values of $\lambda$, leading to underestimation and missing key edges, especially within the first block of the network. Meanwhile, 5-fold CV systematically overestimates some regions of the graph while underestimating others. This visualization underscores the advantage of adaptive tuning in balancing sensitivity and specificity, outperforming fixed global tuning approaches.

\subsection{Estimation}

Results for graph estimation performance are presented in Table~\ref{tab:rmse_off_results}. LARGE achieves off-diagonal RMSE comparable to that of the best-performing benchmark, PC-GLASSO. Consistent with graph selection results, LARGE tends to outperform PC-GLASSO in more challenging DGPs, such as $(n,p)=(300,300)$. We also observe that methods which either adapt to local noise levels (e.g., LARGE and PC-GLASSO), or are tuning-insensitive (e.g., TIGER), generally yield better estimation accuracy than those relying on a single global $\lambda$.

\begin{table}[!htbp]
\centering
\resizebox{\textwidth}{!}{%
\begin{tabular}{l|ccccccc}
\toprule
 &  &  &  & RMSE$_{\mathrm{off}}$  &  &  & 
 \\
Family & LARGE & EBIC & RIC & StARS & 5-fold CV & TIGER & PC-GLASSO \\
\midrule
Band-1 &  &  &  &  &  &  & \\
\hspace{1em}p = 100 &  &  &  &  &  &  & \\
\hspace{2em}n = 300 & 0.46 (0.05) & 1 (0) & 1 (0) & 0.99 (0) & 0.99 (0.03) & 0.84 (0.01) & \textbf{ 0.31 (0.05)} \\
\hspace{2em}n = 500 & 0.32 (0.03) & 0.99 (0) & 0.99 (0) & 0.99 (0) & 0.71 (0.34) & 0.82 (0.01) & \textbf{0.23 (0.11)} \\
\hspace{1em}p = 300 &  &  &  &  &  &  & \\
\hspace{2em}n = 300 & \textbf{0.56 (0.03)} & 1 (0) & 1 (0) & 0.99 (0) & 0.99 (0) & 0.86 (0.01) & 0.77 (0.01) \\
\hspace{2em}n = 500 & 0.38 (0.02) & 0.99 (0) & 1 (0) & 0.99 (0) & 0.99 (0) & 0.83 (0) & \textbf{0.29 (0.02)} \\
\midrule
Band-2 &  &  &  &  &  &  & \\
\hspace{1em}p = 100 &  &  &  &  &  &  & \\
\hspace{2em}n = 300 & \textbf{0.72 (0.04)} & 1 (0) & 1 (0) & 1 (0) & 0.99 (0) & 0.96 (0.01) & 0.82 (0.02) \\
\hspace{2em}n = 500 & 0.59 (0.04) & 1 (0) & 1 (0) & 0.99 (0) & 0.99 (0) & 0.91 (0.01) & \textbf{0.3 (0.03)} \\
\hspace{1em}p = 300 &  &  &  &  &  &  & \\
\hspace{2em}n = 300 & \textbf{0.88 (0.01)} & 1 (0) & 1 (0) & 1 (0) & 1 (0) & 0.98 (0) & 1 (0) \\
\hspace{2em}n = 500 & \textbf{0.70 (0.02)} & 1 (0) & 1 (0) & 0.99 (0) & 0.99 (0) & 0.94 (0) & 0.86 (0.01) \\
\midrule
Block &  &  &  &  &  &  & \\
\hspace{1em}p = 100 &  &  &  &  &  &  & \\
\hspace{2em}n = 300 & 0.90 (0.08) & 1 (0) & 1 (0) & 1 (0) & 1 (0) & 0.97 (0.03) & \textbf{0.89 (0.12)} \\
\hspace{2em}n = 500 & \textbf{0.79 (0.12)} & 1 (0) & 1 (0) & 0.99 (0) & 1 (0) & 0.91 (0.03) & 0.86 (0.1) \\
\hspace{1em}p = 300 &  &  &  &  &  &  & \\
\hspace{2em}n = 300 & \textbf{0.91 (0.03)} & 1 (0) & 1 (0) & 1 (0) & 1 (0) & 0.98 (0.01) & 1 (0.01) \\
\hspace{2em}n = 2000 & \textbf{0.81 (0.04)} & 1 (0) & 1 (0) & 0.99 (0) & 0.99 (0) & 0.93 (0.01) & \textbf{0.81 (0.04)} \\
\midrule
Hub &  &  &  &  &  &  & \\
\hspace{1em}p = 100 &  &  &  &  &  &  & \\
\hspace{2em}n = 300 & \textbf{0.84 (0.06)} & 1 (0) & 1 (0) & 1 (0) & 0.99 (0) & 0.98 (0.01) & 0.92 (0.03) \\
\hspace{2em}n = 500 & 0.77 (0.03) & 1 (0) & 1 (0) & 1 (0) & 0.99 (0.02) & 0.95 (0.01) & \textbf{0.5 (0.03)} \\
\hspace{1em}p = 300 &  &  &  &  &  &  & \\
\hspace{2em}n = 300 & \textbf{0.95 (0.01)} & 1 (0) & 1 (0) & 1 (0) & 1 (0) & 0.99 (0) & 1 (0) \\
\hspace{2em}n = 500 & \textbf{0.88 (0.01)} & 1 (0) & 1 (0) & 1 (0) & 0.99 (0) & 0.97 (0) & 0.93 (0.01) \\
\midrule 
Random &  &  &  &  &  &  & \\
\hspace{1em}p = 100 &  &  &  &  &  &  & \\
\hspace{2em}n = 300 & 0.87 (0.09) & 1 (0) & 1 (0) & 0.99 (0) & 0.99 (0.02) & 0.94 (0.03) & \textbf{0.82 (0.13)} \\
\hspace{2em}n = 500 & 0.76 (0.08) & 1 (0) & 1 (0) & 0.99 (0) & 0.96 (0.05) & 0.85 (0.03) & \textbf{0.74 (0.21)} \\
\hspace{1em}p = 300 &  &  &  &  &  &  & \\
\hspace{2em}n = 300 & \textbf{0.85 (0.02)} & 1 (0) & 1 (0) & 0.99 (0) & 0.99 (0) & 0.94 (0.01) & 0.99 (0.02) \\
\hspace{2em}n = 500 & 0.77 (0.03) & 1 (0) & 1 (0) & 0.99 (0) & 0.93 (0.04) & 0.83 (0.01) & \textbf{0.64 (0.04)} \\
\bottomrule
\end{tabular} }
\caption{\textit{Mean  off-diagonal RMSE of LARGE and benchmark methods across all simulation settings.} Bolded values indicate the best performance in each setting. Standard
deviations are reported in parentheses. Results are averaged over 50 replications.}
\label{tab:rmse_off_results}
\end{table}

\subsection{Convergence} 

We monitor the convergence of LARGE using the relative Frobenius norm:
\begin{equation}
\frac{\| W_{\text{new}} - W_{\text{old}} \|_F}{\| W_{\text{old}} \|_F} < e.
\end{equation}
In practice, \textsc{LARGE} typically converges within 20 iterations. We observe one instance of non-convergence in the band-2 scenario and three instances in the hub scenario. To ensure convergence in these more challenging data-generating processes, we use $e = 0.005$ as the convergence threshold for $ p = 100$, and relax to $e = 0.05$ for $p = 300$. While this adjustment enables successful convergence, it highlights a practical limitation: users may need to relax the convergence threshold depending on the problem's dimension and structural complexity.

\section{Data Analysis}\label{sec:realdata}

Functional connectivity (FC) refers to the temporal dependency patterns between regional blood-oxygen-level-dependent (BOLD) signals measured via functional magnetic resonance imaging (fMRI). To estimate FC, the brain is first parcellated into anatomically defined regions. For each region, the fMRI scans record the BOLD signal over time, which reflects the level of neural activity at each time point. FC has been associated with cognitive functioning and is often used to predict cognitive performance in healthy individuals \citep{seeley2007dissociable, van2009efficiency}. 

Traditionally, FC has been computed using Pearson correlation to quantify the linear associations between pairs of regions \citep{Li2019GSR, He2020DNN, dhamala2021distinct}. However, alternative measures, such as the empirical covariance and Tikhonov precision matrix, have also been explored \citep{pervaiz2020tikinov}. In this analysis, we apply LARGE to estimate and visualize the FC network as a GGM. The data set used here is a publicly available, high-resolution, preprocessed fMRI data from the Human Connectome Project (HCP) - Young Adult S1200 release \citep{VanEssen2013HCP}. We consider the time series data of 1003 healthy adults observed over same time horizon across four complete resting state fMRI runs, each scan with $n = 1200$ time points.

\underline{\textit{Preprocessing:}} MRIs were acquired on a Siemens Skyra 3T scanner at Washington University in St. Louis. Each subject underwent four gradient-echo EPI resting-state functional MRIs (rsfMRI), with TR = 720 ms, TE = 33.1 ms, 2.0 mm isotropic voxels, FoV = 208~$\times$~180~mm\textsuperscript{2}, flip angle = 52\textdegree, matrix = 104~$\times$~90, 72 slices. Two 15-minute rsfMRI runs were acquired at each of two sessions. In total, the data consisted of 1200 time points per run, for a total of 4800 time points per subject across four scans. Each run of each subject’s rsfMRI was preprocessed by the HCP consortium \citep{Smith2011fmri}. The data were minimally preprocessed \citep{glasser2013minimal} and had artifacts removed using ICA+FIX \citep{griffanti2014ica, salimi2014automatic}. We applied a standard post-processing pipeline, as described in \cite{jamison2024Krakencoder}, to identify motion and global signal outlier timepoints, regress out tissue-specific and motion-related nuisance time series, and apply temporal filtering. Outlier timepoints were excluded from both nuisance regression and temporal filtering. Gray matter was parcellated into 86 cortical, subcortical, and cerebellar regions using FreeSurfer \citep{fischl1999cortical}, and regional time series were calculated as the mean of the time series in each voxel of a given region. 

\begin{figure}[ht]
    \centering
    \includegraphics[width=0.7\textwidth]{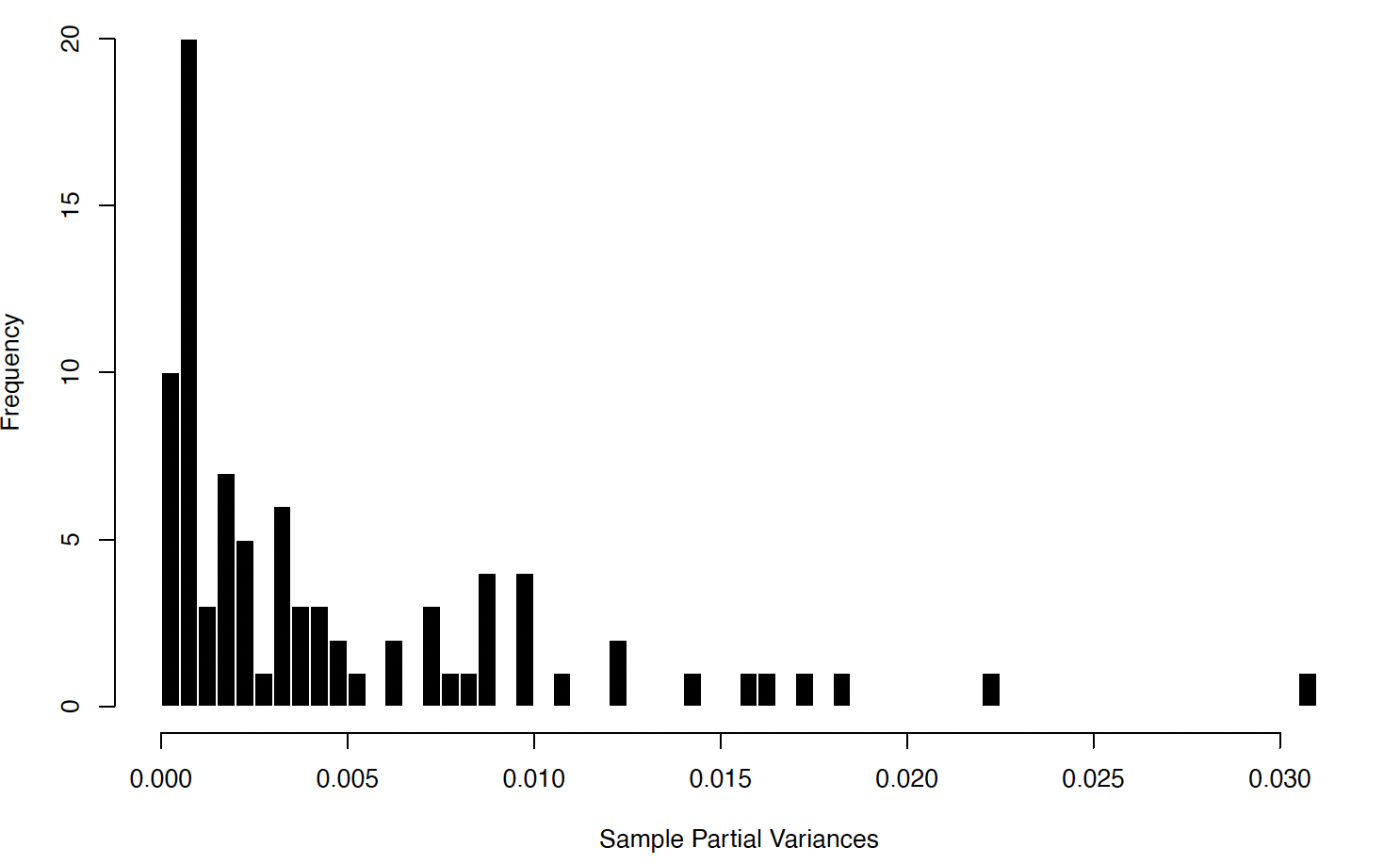}
    \caption{\textit{Distribution of sample partial variances of across 86 brain regions with $n = 1200$ time points from one fMRI scan.} }
    \label{fig:FC_partial_vars}
\end{figure}

\underline{\textit{Results:}} For a randomly selected subject, we begin by plotting the histogram of the sample partial variances from the first fMRI scan ($n = 1200$ time points) by inverting the sample covariance matrix. As shown in Figure \ref{fig:FC_partial_vars}, the partial variances  exhibit considerable variability, suggesting that we might benefit from using locally adaptive nodewise penalties when estimating FC.

\begin{figure}[ht]
    \centering
    \includegraphics[width=1\textwidth]{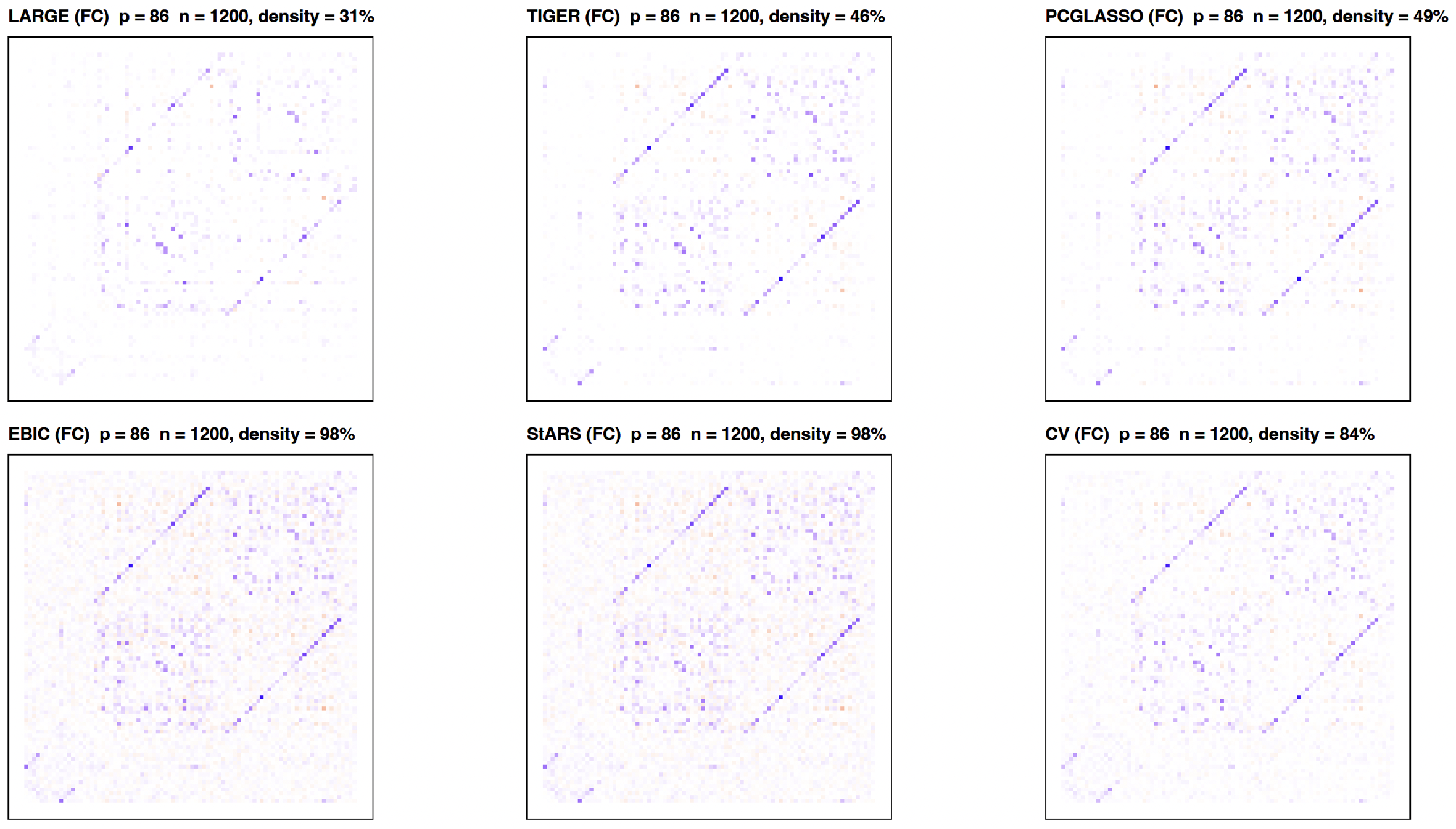}
    \caption{\textit{FC networks estimated by LARGE and benchmark methods for a single individual, visualized on the correlation scale.} All methods recover known biological patterns of brain connectivity.}
    \label{fig:FC}
\end{figure}

To visualize and interpret the FC network, we estimate the precision matrix from the fMRI time series of 86 brain regions for each scan using LARGE and benchmark methods. Since resulting partial variances are relatively small, we apply penalization to the diagonal entries of the precision matrix to improve numerical stability. We then average the resulting precision matrices across the four scans and display the final network in Figure~\ref{fig:FC}. We exclude RIC as benchmark from this comparison, as it yields overly sparse estimates that lack interpretability.

All methods appear effective at capturing the underlying sparsity structure. Their estimated precision matrices preserve known physiological patterns, most notably, strong connections between bilateral homologues (i.e., corresponding regions in the left and right hemispheres), which are well-established in the literature \citep{zuo2010growing}. In addition, we observe a modular block structure separating the left and right cortical hemispheres, as well as a distinct subcortical cluster \citep{power2011functional}. Among these methods, LARGE produces the sparsest solution (density = 31\%), which tends to suppress weaker connections and may enhance interpretability. In contrast, CV, EBIC, and StARS yield denser networks (density $>$ 80\%) that may introduce additional noise.

\section{Conclusion}\label{sec:conclusion}

In this paper, we introduce an adaptive framework for estimating sparse precision matrices that learns nodewise, scale-appropriate \(\ell_1\)-penalties for each column of the inverse covariance matrix. Our method augments a block coordinate descent algorithm with an inner loop that dynamically tracks nodewise noise variance estimates, which in turn are used to derive locally adaptive tuning parameters. To further enhance edge recovery, we incorporate a variable selection strategy that guides the Lasso path toward 
pairs of nodes with high marginal correlation, enabling early detection of true edges in the graph. 
Overall, our approach marks a conceptual shift from globally tuned, scale-sensitive regularization $\lambda$ to a more interpretable, hypothesis-testing-based control of sparsity via a user-specified significance level $\alpha$.

The promising empirical performance of our method raises several exciting research directions. On the algorithmic side, it remains an open question how to assess the suitability of \(\ell_1\)-based methods, e.g.,  through goodness-of-fit checks that evaluate whether the true graph is approximately sparse. On the computational side, the algorithm could benefit from more efficient implementations of key steps such as sorting variables by the standard deviation of partial residuals or performing sequential F-tests for variable selection. On the theoretical side, a rigorous convergence analysis of the block coordinate descent algorithm with a biconvex objective in the inner loop would further clarify the scope and guarantees of the proposed method. We leave these important questions for future work.

\section*{Acknowledgements}

SB acknowledges partial support from NIH award R01GM135926, NSF awards DMS-1812128, DMS-2210675, and CAREER DMS-2239102. HN and SB are grateful for the insightful input from Dr. Amy Kuceyeski and Dr. Keith Jamison on the analysis and interpretation of the neuroscience application.

\newpage
\bibliographystyle{abbrvnat}
\bibliography{biblio}

\newpage
\appendix
\newpage
\section{Appendix}\label{sec:appendix}

\subsection{LARGE with Diagonal Penalization}

In this section, we present an alternative formulation and algorithm of LARGE where both diagonal and off-diagonal entries of $\Theta$ are penalized. 
\begin{equation}
    \hat{\Theta} := 
    \arg\min_{\Theta \in \mathbb{S}_{++}^p} \left\{-  \log \det({\Theta}) + \mathrm{tr}({S} {\Theta}) + \sum_{j=1}^p \lambda_j \|{\Theta_{.j}}\|_{1} \right\},
    \label{eq:large_model_penalize} 
\end{equation}
where ${\|\Theta_{.j}}\|_{1} = \sum_{k=1}^p |\Theta_{kj}|$.

\begin{algorithm}[ht]
\caption{\textbf{LARGE with Diagonal Penalization}}
\label{alg:large_alg_penalized}
\begin{algorithmic}[1]
\Statex \textbf{Input:} Data matrix \(\mathcal{X} \in \mathbb{R}^{n \times p}\), Sequential F-test significance level $\alpha$
\State For $j=1,\ldots, p$, initialize \(\hat{\sigma}^2_j \gets \mathrm{Var}(\mathcal{X}_{\cdot j})\),  \(\lambda_{0j} \gets \frac{1}{\mathrm{var}(\mathcal{X}_{\cdot j})} \max_{k \ne j} \left| \frac{1}{2n} \langle \mathcal{X}_{\cdot j}, \mathcal{X}_{\cdot k} \rangle \right|\)
\State Let $S \gets \mathrm{Cov}(\mathcal{X})$; initialize $W \gets S$

\Repeat \text{ for \( j = 1, 2, \ldots, p \):}
\State (a) Partition $
W = \begin{bmatrix}
W_{11} & w_{12} \\
w_{12}^\top & w_{22}
\end{bmatrix}, \quad
S = \begin{bmatrix}
S_{11} & s_{12} \\
s_{12}^\top & s_{22}
\end{bmatrix}.
$ 

    \State (b) Solve $W_{11} \beta - s_{12} + \lambda_{j} \cdot \text{sign}(\beta) = 0$ using Autotune Lasso:
    \State Let $V \gets W_{11}$; initialize $\hat{\beta} \gets 0$
    \State Initialize $\texttt{Predictor.Ranking}_{j} \gets \{1, \ldots, p-1\}$, $\texttt{Support.Set}_{j} \gets \emptyset$, $\texttt{sigma.update.flag}_{j}  \gets \texttt{TRUE}$


    \While{$\texttt{error} \geq \texttt{error.tolerance}$}
    \State $\texttt{Support.Set}_j^{(\text{old})} \gets \texttt{Support.Set}_j,\quad \hat{\beta}^{(\text{old})} \gets \hat{\beta}$

        \For{$k$ in $\texttt{Predictor.Ranking}_{j}$}
            \State \(\hat{\beta}_k \gets \texttt{Soft.Threshold}_{\textcolor{black}{\lambda_j = \hat{\sigma}^2_j \cdot \lambda_{0j}}} \left(s_{12k} - \sum_{l \ne k} V_{kl} \cdot \hat{\beta}_l \right) \big/ (V_{kk} + \lambda_k)\)
        \EndFor
        \If{$\texttt{sigma.update.flag}_j == \texttt{TRUE}$}
        \State $\texttt{Predictor.Ranking}_j \gets$ Rank \(k\) in decreasing order of:
        \[
        \begin{cases}
        \textcolor{black}{\left| \mathrm{Cor}(\mathcal{X}_{\cdot j}, \mathcal{X}_{\cdot k}) \right|}, & \text{if first iteration of line 8} \\
        \mathrm{SD} \text{ of } r_k = \mathcal{X}_{\cdot j} - \sum_{l \notin \{ j, k \}} \mathcal{X}_{\cdot l} \hat{\beta}_l, & \text{otherwise}
        \end{cases}
        \]
        \State Fit a linear model using sequential F-tests for variable selection with cutoff $=F_{\alpha; 1, n-i}$
        \State \(\texttt{Support.Set}_j \gets  \{ k : \mathcal{X}_{\cdot k} \text{ selected into the linear model} \}\)
        \State \textcolor{black}{\(\hat{\sigma}^2_j \gets \frac{\sum_{i=1}^n (X_{ij} - \hat{X}_{ij})^2}{n - |\texttt{Support.Set}_j|}\)}
        \If{$\texttt{Support.Set}_j \subseteq \texttt{Support.Set}_j^{(\text{old})}$}
            \State $\texttt{sigma.update.flag}_j \gets \texttt{FALSE}$
        \EndIf
     \EndIf
    \EndWhile
    \State (c) \( w_{12} \gets \large(W_{11} + \mathrm{diag(\lambda_{-j})I} \large) \hat{\beta} \) with symmetrization
\Until{convergence over all \(j\)}
\State For $i = 1, \ldots, p$, \(\hat{\Theta}_{jj} \gets (\hat{\sigma}_j + \lambda_j)^{-2}\) and \(\hat{\Theta}_{j,-j} \gets -\hat{\beta} \hat{\Theta}_{jj}\) with symmetrization
\Statex \textbf{Output:} \(\hat{\Theta}\), $(\hat{\sigma}_j^2, \lambda_j)$ for $j=1,\ldots, p$
\end{algorithmic}
\end{algorithm}

\end{document}